\documentclass{acm_proc_article-sp}

\usepackage{algorithm}
\usepackage{algpseudocode}
\usepackage{amsbsy}
\usepackage{amsmath}
\usepackage{color}
\usepackage{enumerate}
\usepackage{float}
\usepackage{graphicx}
\usepackage[breaklinks]{hyperref}
\usepackage{breakurl}
\usepackage[final]{listings}
\usepackage{multirow}
\usepackage{rotating}
\usepackage[caption=false]{subfig}
\usepackage{tikz}
\usepackage{verbatim}
\usepackage{xspace}

\newcommand{\ie}{{\em i.e.}\xspace}
\newcommand{\eg}{{\em e.g.}\xspace}
\newcommand{\etal}{{\em et al.}\xspace}

\newcommand{\Or}{\textbf{or}}

\algnewcommand\algorithmicswitch{\textbf{switch}}
\algnewcommand\algorithmiccase{\textbf{case}}
\algnewcommand\algorithmicassert{\texttt{assert}}
\algnewcommand\Assert[1]{\State \algorithmicassert(#1)}%
\algdef{SE}[SWITCH]{Switch}{EndSwitch}[1]{\algorithmicswitch\ #1\ \algorithmicdo}{\algorithmicend\ \algorithmicswitch}%
\algdef{SE}[CASE]{Case}{EndCase}[1]{\algorithmiccase\ #1}{\algorithmicend\ \algorithmiccase}%
\algtext*{EndSwitch}%
\algtext*{EndCase}%

\newcommand{\PRAGMATIC}{PRAgMaTIc\xspace}

\newcommand{\INTELXEON}{Intel\textsuperscript\textregistered\allowbreak 
Xeon\textsuperscript\textregistered\xspace}

\newcommand{\AMDOPTERON}{AMD Opteron\texttrademark\xspace}
\newcommand{\RHEL}{Red Hat\textsuperscript\textregistered\allowbreak Enterprise 
Linux\textsuperscript\textregistered\xspace}

\definecolor{listinggray}{gray}{0.9}
\definecolor{lbcolor}{rgb}{0.9,0.9,0.9}
\floatstyle{plain}
\newfloat{code}{h}{}
\floatname{code}{Code Snippet}

\begin{document}

\title{Thread Parallelism for Highly Irregular Computation in Anisotropic Mesh 
Adaptation}

\numberofauthors{2}
\author{
\alignauthor
Georgios Rokos\\
       \affaddr{Imperial College London}\\
       \affaddr{London, United Kingdom}\\
       \email{georgios.rokos09@imperial.ac.uk}
\alignauthor
Gerard J. Gorman\\
       \affaddr{Imperial College London}\\
       \affaddr{London, United Kingdom}\\
       \email{g.gorman@imperial.ac.uk}
\and
\alignauthor
Kristian Ejlebjerg Jensen\\
       \affaddr{Imperial College London}\\
       \affaddr{London, United Kingdom}\\
       \email{kristianejlebjerg@gmail.com}
\alignauthor
Paul H. J. Kelly\\
       \affaddr{Imperial College London}\\
       \affaddr{London, United Kingdom}\\
       \email{p.kelly@imperial.ac.uk}
}

\maketitle
\begin{abstract}
Thread-level parallelism in irregular applications with mutable data 
dependencies presents challenges because the underlying data is extensively 
modified during execution of the algorithm and a high degree of parallelism 
must be realized while keeping the code race-free. In this article we describe 
a methodology for exploiting thread parallelism for a class of graph-mutating 
worklist algorithms, which guarantees safe parallel execution via processing in 
rounds of independent sets and using a deferred update strategy to commit 
changes in the underlying data structures. Scalability is assisted by atomic 
fetch-and-add operations to create worklists and work-stealing to balance the 
shared-memory workload. This work is motivated by mesh adaptation algorithms, 
for which we show a parallel efficiency of 60\% and 50\% on \INTELXEON Sandy 
Bridge and \AMDOPTERON Magny-Cours systems, respectively, using these 
techniques.
\end{abstract}


\section{Introduction}
Finite element/volume methods (FEM/FVM) are commonly used in the numerical 
solution of partial differential equations (PDEs). Unstructured meshes, where 
the spatial domain has been discretised into simplices (\ie triangles in 2D, 
tetrahedra in 3D), are of particular interest in applications where the 
geometric domain is complex and structured meshes are not practical. Simplices 
are well suited to varying mesh resolution throughout the domain, allowing for 
local coarsening and refinement of the mesh without hanging nodes. On the other 
hand, this flexibility introduces complications of its own, such as management 
of mesh quality and computational overheads arising from indirect addressing.

Computational mesh resolution is often the limiting factor in simulation 
accuracy. Being able to accurately resolve physical processes at the small 
scale coupled with larger scale dynamics is key to improving the fidelity of 
numerical models across a wide range of applications (\eg \cite{pain2005three, 
southern2011parallel}). A difficulty with mesh-based modelling is that the mesh 
is generated before the solution is known, however, the local solution error is 
related to the local mesh resolution. Overly coarse meshes lead to low accuracy 
whereas over-refined meshes can greatly increase the computational cost.

Mesh adaptation methods provide an important means to minimise computational 
cost while still achieving the required accuracy \cite{M.D.Piggott11282009, 
li20053d}. In order to use mesh adaptation within a simulation, the application 
code requires a method to estimate the local solution error. Given an error 
estimate it is then possible to compute a solution to a specified error 
tolerance while using the minimum resolution everywhere in the domain and 
maintaining element quality constraints.

Previous work has described how adaptive mesh methods can be parallelised for 
distributed-memory systems using MPI (\eg \cite{li20053d, 
Freitag98thescalability}). However, there is a continuous trend towards an 
increasing number of cores per compute node in the world's most powerful 
supercomputers and it is assumed that the nodes of a future exascale system 
will each contain thousands of cores \cite{dongarra2011exascale}. Therefore, it 
is important that algorithms are developed with very high levels of parallelism 
and using thread-parallel programming models, such as OpenMP \cite{660313}, 
that exploit the memory hierarchy. However, irregular applications are hard to 
parallelise effectively on shared-memory architectures for reasons described in 
\cite{doi:10.1142/S0129626407002843}.

In this article we take a fresh look at anisotropic adaptive mesh methods and 
parallelise them using new scalable techniques suitable for modern multicore 
and manycore architectures. These concepts have been implemented in the open 
source framework \PRAGMATIC (Parallel anisotRopic Adaptive Mesh 
ToolkIt)\footnote{http://meshadaptation.github.io/}. The remainder of the paper 
is laid out as follows: \S \ref{sect:overview} gives an overview of the mesh 
adaptation procedure; \S \ref{sect:methodology} describes the new irregular 
compute methodology used to parallelise the adaptive algorithms; and \S 
\ref{sect:benchmark} illustrates how well our framework performs in 2D and 3D 
benchmarks. We conclude the paper in \S \ref{sect:conclusion}.

\section{Mesh Adaptivity Background}
\label{sect:overview}
In this section we give an overview of anisotropic mesh adaptation, focusing on 
the element quality as defined by an error metric and the adaptation kernels 
which iteratively improve local mesh quality as measured by the worst local 
element.

\subsection{Error control}
\label{subsect:error_control}
Solution discretisation errors are closely related to the size and the shape of 
the elements. However, in general meshes are generated using {\em a priori} 
information about the problem under consideration when the solution error 
estimates are not yet available. This may be problematic because (a) solution 
errors may be unacceptably high and (b) parts of the solution may be 
over-resolved, thereby incurring unnecessary computational expense. A solution 
to this is to compute appropriate local error estimates and use them to 
dynamically control the local mesh resolution at runtime. In the most general 
case this is a metric tensor field so that the resolution requirements can be 
specified anisotropically; for a review of the procedure see 
\cite{frey2005anisotropic}.

\subsection{Element quality}
\label{subsect:element_quality}
As discretisation errors are dependent upon element shape as well as size, a 
number of measures of element quality have been proposed, out of which, in the 
work described here, we use the quality functionals by Vasilevskii \etal for 
triangles \cite{vasilevskii1999adaptive} and tetrahedra 
\cite{agouzal-1999-adaptive}, which indicate that the ideal element is an 
equilateral triangle/tetrahedron with edges of unit length measured in metric 
space.

\subsection{Overall adaptation procedure}
\label{subsect:overall_procedure}
The mesh is adapted through a series of local operations: edge collapse, edge 
refinement, element-edge swaps and vertex smoothing. While the first two of 
these operations control the local resolution, the latter two are used to 
improve the element quality. Algorithm \ref{alg:general} gives a high level 
view of the anisotropic mesh adaptation procedure as described by Li \etal 
\cite{li20053d}. The inputs are $\mathcal{M}$, the piecewise linear mesh from 
the modelling software, and $\mathcal{S}$, the node-wise metric tensor field 
which specifies anisotropically the local mesh resolution requirements. The 
process involves the iterative application of coarsening, swapping and 
refinement to optimise the resolution and quality of the mesh. The loop 
terminates once the mesh optimisation algorithm converges or after a maximum 
number of iterations has been reached. Finally, mesh quality is fine-tuned 
using some vertex smoothing algorithm, which aims primarily at improving the 
worst-element quality. Smoothing is left out of the main loop because it is an 
expensive operation and it is found empirically that it is efficient to 
fine-tune the worst-element quality once mesh topology has been fixed.

\begin{algorithm}
  \caption{Mesh optimisation procedure.}
  \label{alg:general}
  \begin{algorithmic}
	\State Inputs: $\mathcal{M}$, $\mathcal{S}$.
	\Repeat
	  \State $(\mathcal{M}^*, \mathcal{S}^*) \gets coarsen(\mathcal{M}^*$, 
	  $\mathcal{S}^*)$
	  \State $(\mathcal{M}^*, \mathcal{S}^*) \gets swap(\mathcal{M}^*$, 
	  $\mathcal{S}^*)$
	  \State $(\mathcal{M}^*, \mathcal{S}^*) \gets refine(\mathcal{M}^*$, 
	  $\mathcal{S}^*)$
	\Until{(max. number of iterations \Or ~convergence)}
    \State $(\mathcal{M}^*, \mathcal{S}^*) \gets smooth(\mathcal{M}^*$, 
    $\mathcal{S}^*)$
    \State \Return $\mathcal{M}^*$
  \end{algorithmic}
\end{algorithm}

\subsection{Adaptation kernels}
\label{subsect:kernels}
\begin{figure}
\centering
\epsfig{file=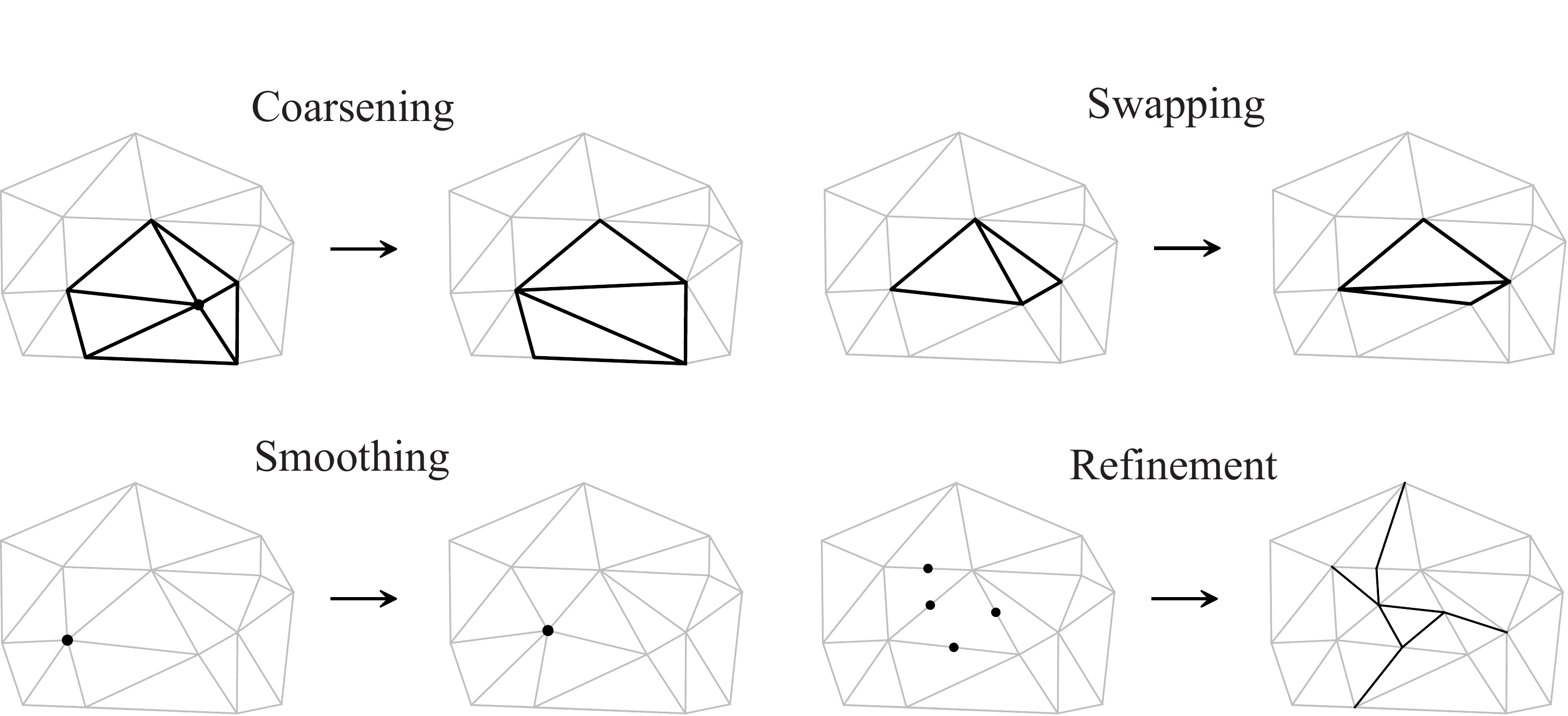, width=\linewidth}
\caption{Examples of the four adaptive kernels.}
\label{fig:anisoadapt}
\end{figure}

A brief description of the four mesh optimisation kernels follows. Figure 
\ref{fig:anisoadapt} shows 2D examples to demonstrate what each kernel does to 
the local mesh patch, but the same operations are applied in an identical 
manner in 3D. For more details on the adaptive algorithms the reader is 
referred to the publications by Li \etal \cite{li20053d} (coarsening, 
refinement, swapping) and Freitag \etal \cite{freitag1996comparisonw, 
NME:NME341} (smoothing).

\subsubsection{Coarsening}
\label{subsubsect:coarsening}
Coarsening is the process of lowering mesh resolution locally by collapsing an 
edge to a single vertex, thereby removing all elements that contain this edge, 
leading to a reduction in the computational cost.


\subsubsection{Refinement}
\label{subsubsect:refinement}
Refinement is the process of increasing mesh resolution locally by (a) 
splitting of edges which are longer than desired (as indicated by the error 
estimation) and (b) subsequent division of elements using refinement templates 
\cite{NME:NME741}.


\subsubsection{Swapping}
\label{subsubsect:swapping}
Swapping is done in the form of flipping an edge shared by two elements, 
considering the quality of the swapped elements - if the minimum quality is 
improved then the original mesh elements are replaced with the edge-swapped 
elements.


\subsubsection{Smoothing}
\label{subsubsect:smoothing}
The kernel of vertex smoothing relocates a central vertex so that the local 
mesh quality is increased. Common heuristic methods are the quality-constrained 
Laplacian smoothing \cite{freitag1996comparisonw} and the more expensive 
optimisation-based smoothing \cite{NME:NME341}.


\subsubsection{Propagation}
\label{subsubsect:propagation}
The operations of coarsening, swapping and smoothing often need to be 
propagated to the local mesh neighbourhood. When a kernel is applied onto an 
edge/vertex, neighbouring edges/vertices need to be reconsidered for processing 
because the topological/geometrical changes that occurred might give rise to 
new configurations of better quality. Therefore, these adaptive algorithms keep 
sweeping over the mesh until no further changes are made.

\section{Irregular Compute Method}
\label{sect:methodology}
To allow fine grained parallelisation of mesh adaptation we based our 
methodology on graph colouring, following a proposal by Freitag \etal 
\cite{Freitag98thescalability}. However, while this approach avoids updates 
being applied concurrently to the same neighbourhood, data writes will still 
incur significant lock and synchronisation overheads. For this reason we 
incorporate a deferred update strategy, described below, to minimise 
synchronisations and allow parallel writes. Additionally, we make use of atomic 
operations to create parallel worklists in a synchronisation-free fashion and, 
finally, try to balance the workload among threads using work-stealing 
\cite{Blumofe:1994:SMC:1398518.1398998}.

\subsection{Hazards}
\label{subsect:hazards}
There are two types of hazards when running mesh optimisation algorithms in 
parallel: {\em topological hazards}; and {\em data races}. The former refers to 
the situation where an adaptive operation results in invalid or non-conforming 
edges and elements. For example in coarsening, if some vertex $V_B$ collapses 
onto another vertex $V_A$, then $V_A$ cannot collapse onto some other vertex at 
the same time. Data races can occur when two threads try to update the same 
adjacency list of a vertex concurrently. For example in coarsening, two 
neighbours of some vertex $V_A$ can collapse onto $V_A$ concurrently, then 
$V_A$'s adjacency list has to be updated to reflect the changes made by the 
coarsening operations. Concurrent access to $V_A$'s adjacency list may lead to 
race conditions.

\subsection{Colouring}
\label{subsect:colouring}
Topological hazards for all adaptive algorithms are avoided by colouring a 
graph whose nodes are defined by the mesh vertices and edges are defined by the 
mesh edges. The adaptive algorithm then processes the mesh in batches of 
independent sets. The fact that topological changes are made to the mesh means 
that colouring is invalidated frequently and the mesh has to be re-coloured 
before proceeding to the next iteration of the adaptive algorithm. Therefore, 
we need to use a fast and scalable colouring algorithm (see 
\cite{doi:1505.04086}).

\subsection{Deferred Update}
\label{subsect:deferred_update}
Colouring does not eliminate data races when updating adjacency lists. A 
2-distance colouring was not considered here as it is expensive and increases 
the chromatic number, effectively reducing the exposed parallelism. Instead, in 
a shared-memory environment with $N$ threads, each thread allocates a private 
collection of $N$ lists. When the adjacency list of some vertex $V_i$ has to be 
updated, the thread executing the adaptive kernel does not commit the update 
immediately; instead, it pushes the operation back into the list for thread 
$tid=ID(V_i)\%N$, where $ID(V_i)$ is the integer identifier of $V_i$. After 
processing an independent set and before proceeding to the next one, every 
thread $T_i$ visits the private collections of all threads, locates the list 
reserved for $T_i$ and commits the updates stored there. This way, it is 
guaranteed that one and only thread will update the adjacency list of any given 
vertex. We call this technique the {\em deferred update}. Code Snippet 
\ref{code:deferred_update} demonstrates a typical usage scenario. An important 
advantage of this strategy is that we always read the most up-to-date data when 
executing an adaptive kernel (as if we used an ``as we go'' write-back scheme), 
eliminating the risk of mesh data corruption in coarsening, refinement and 
swapping and having a faster-converging Gauss-Seidel-style iteration process in 
smoothing.

\begin{code}
\caption{Example of the deferred update scheme.}
\label{code:deferred_update}
\begin{lstlisting}
typedef std::vector<Updates> DefUpdList;
int N = omp_get_max_threads();

// N collections of deferred-update lists
std::vector< std::vector<DefUpdList> > defUpd(N);

#pragma omp parallel
{
  int tid = omp_get_thread_num();
  // Allocate one list for each thread.
  defUpd[tid].resize(N);
  
  // Process the independent set in parallel
  #pragma omp for
  for(int i=0; i<nVerticesInSet; ++i){
    update = execute_kernel(i);
    // To be committed by thread i%N.
    defUpd[tid][i%N].push_back(update);
  }
  
  // Commit updates tid is responsible for.
  for(int i=0; i<N; ++i)
    commit_all_updates(defUpd[i][tid]);
  
  // Proceed to the next independent set...
}
\end{lstlisting}
\end{code}

\subsection{Worklists and Atomic-Capture}
\label{subsect:worklists}
There are many cases where it is necessary to create a worklist of items which 
need to be processed, \eg for propagation of adaptive operations. New work 
items generated locally by a thread need to be accumulated into a global 
worklist over which the next invocation of the adaptive kernel will iterate. 
The classic approach based on prefix sums \cite{BlellochTR90} requires thread 
synchronisation and was found limiting in terms of scalability. A better method 
is based on atomic fetch-and-add on a global integer which stores the size of 
the worklist needed so far. Every thread increments this integer atomically 
while caching the old value. This way, the thread knows where to copy its 
private data and increments the integer by the size of this data, so the next 
thread to access the integer knows in turn where to copy its private data. An 
example of using this technique via OpenMP's atomic-capture clause 
\cite{openmp40} is given in Code Snippet \ref{code:atomic_capture}, where it is 
shown that no thread synchronisation is needed to generate the global worklist 
(note the \verb=nowait= clause). The overhead/spinlock associated with 
atomic-capture operations was found to be insignificant.

\begin{code}
\caption{Creating a worklist using atomic-capture.}
\label{code:atomic_capture}
\begin{lstlisting}
int worklistSize = 0;
std::vector<Item> globalWorklist(prealloc_size);

#pragma omp parallel
{
  std::vector<Item> private_list;

  #pragma omp for nowait
  for(all items which need to be processed){
    new_item = do_some_work();
    private_list.push_back(new_item);
  } // No need to synchronise at end of loop.
  
  int idx;
  #pragma omp atomic capture
  {
    idx = worklistSize;
    worklistSize += private_list.size();
  }
  
  memcpy(&globalWorklist[idx], &private_list[0],
             private_list.size() * sizeof(Item));
}
\end{lstlisting}
\end{code}

\subsection{Work-stealing}
\label{subsect:work_stealing}
Work-stealing \cite{Blumofe:1994:SMC:1398518.1398998} is a sophisticated 
technique aiming at balancing workload among threads while keeping scheduling 
overhead as low as possible. For-loop scheduling strategies provided by the 
OpenMP runtime system were found to be inadequate, either incurring significant 
scheduling overhead or leading to load imbalances. As of version 4.0, OpenMP 
does not support work-stealing for parallel for-loops so we created a novel 
scheduler \cite{doi:1505.04134} which differs from other proposals in two ways: 
it engages a heuristic to help the thief find a suitable victim to steal from; 
and uses POSIX signals/interrupts to accomplish stealing in an efficient manner.

\section{Experimental Results}
\label{sect:benchmark}
We will show adaptivity results for viscous fingering in 2D and structural 
compliance minimisation in 2D and 3D, followed by performance evaluation.

\subsection{Viscous Fingering}
\begin{figure}
\begin{minipage}{0.49\linewidth}
\centering
\epsfig{file=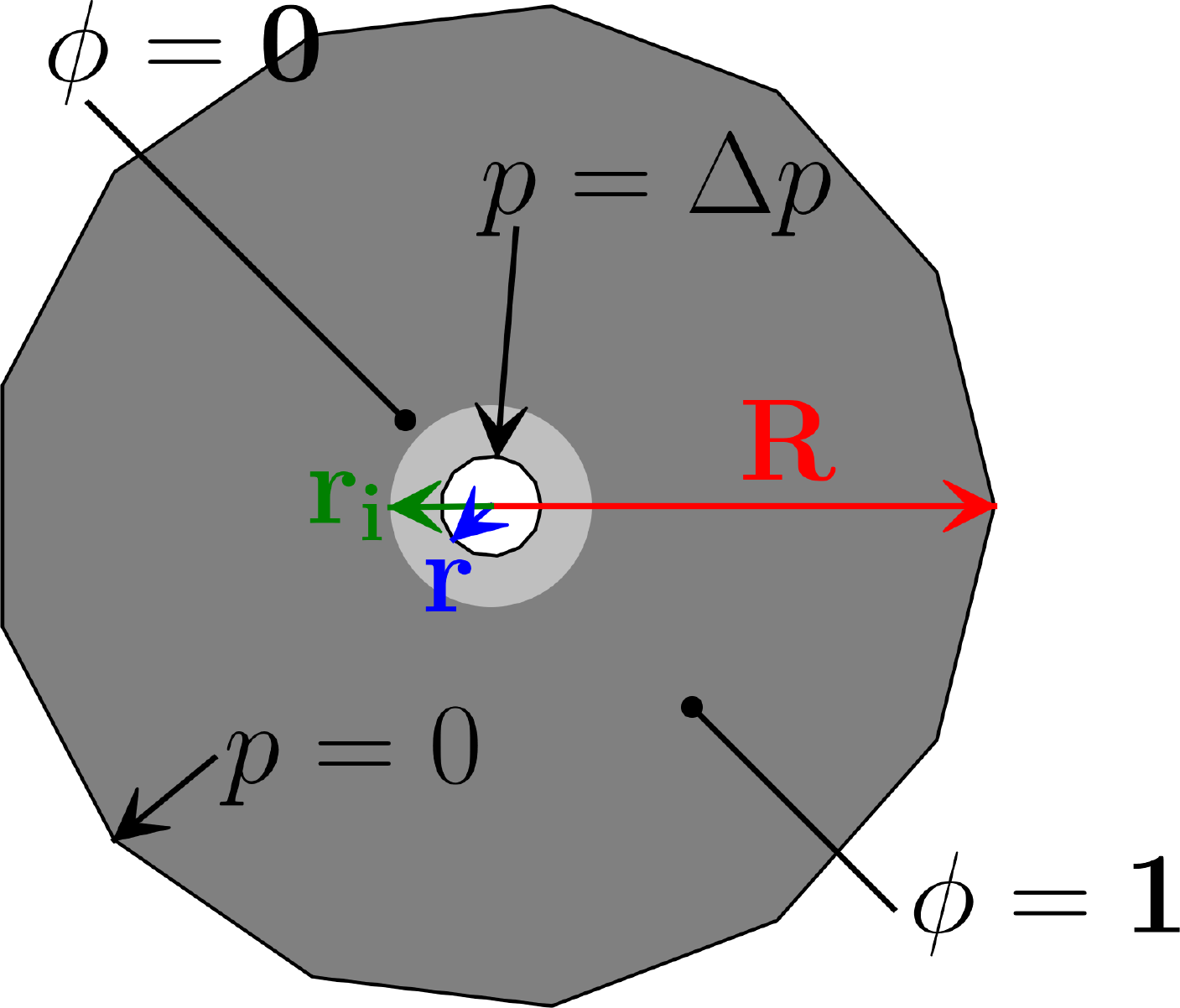, width=\linewidth}
\end{minipage}
\begin{minipage}{0.49\linewidth}
\centering
\epsfig{file=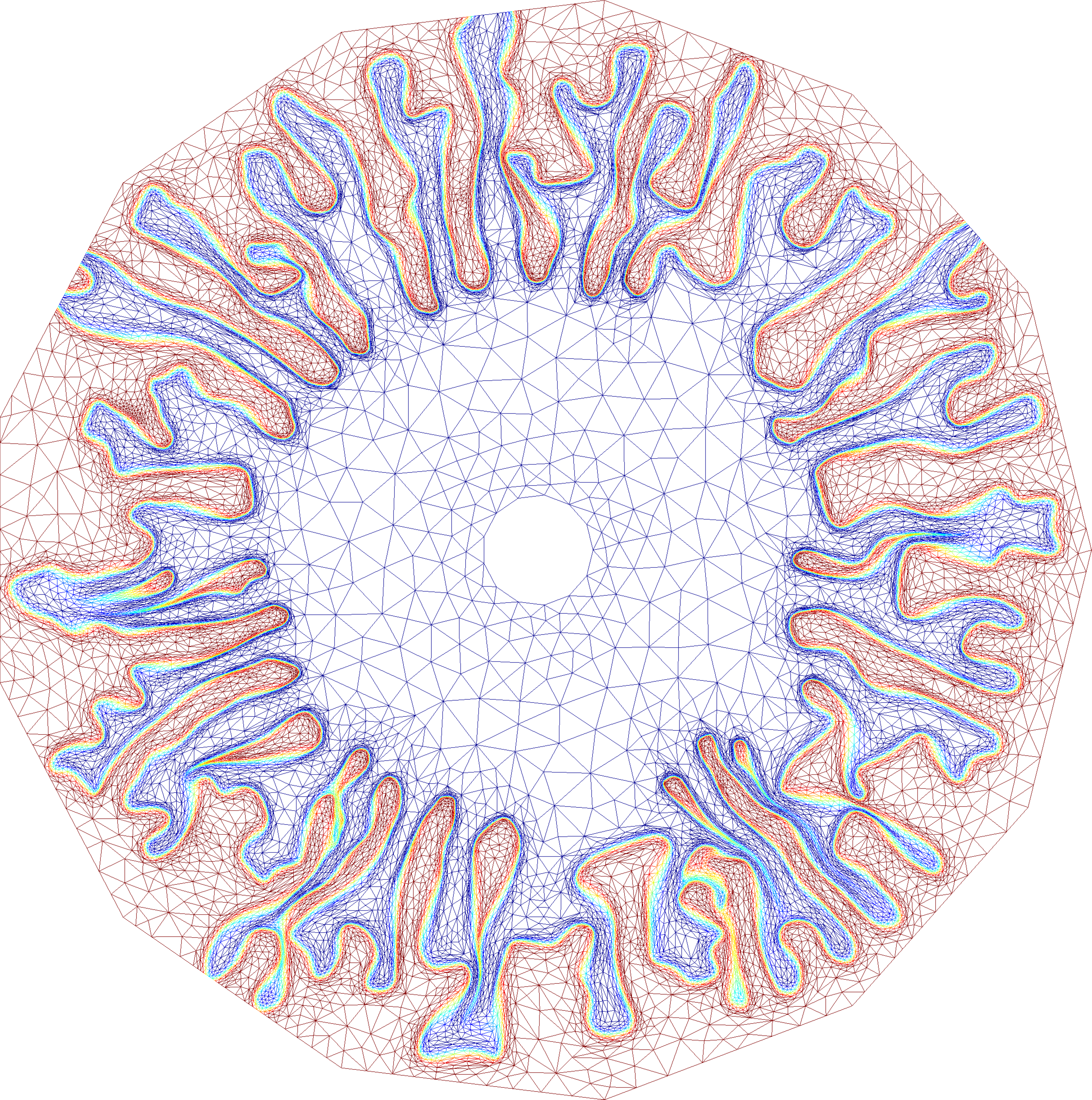, width=\linewidth}
\end{minipage}
\caption{The initial condition for the viscous fingering (left) and a snapshot 
of a simulation (right).}
\label{fig:vf}
\end{figure}

Viscous fingering is a limiting process for enhanced oil recovery technologies. 
It happens whenever one fluid displaces another with a higher viscosity 
\cite{pramanik2012miscible}, typically in a porous media. A typical setup and 
simulation is shown in Figure \ref{fig:vf}, with the blue fluid having a 
viscosity $e^2$ times lower than the red fluid. The initial saturation is 
unperturbed and it is thus the length scale of the initial mesh that triggers 
the instability. Mesh adaptation is driven by the Hessian of the pressure 
combined with the Hessian of the saturation $\phi$ \cite{chen2007optimal}.

\subsection{Structural Optimisation}
\begin{figure}
\begin{minipage}{0.49\linewidth}
\centering
\epsfig{file=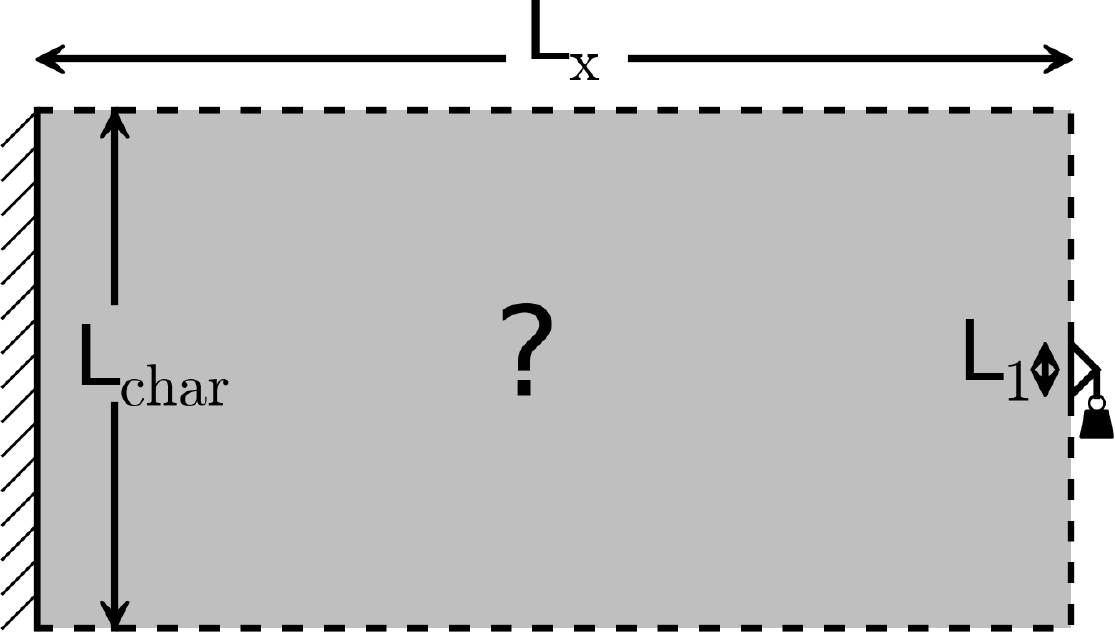, width=\linewidth}
\end{minipage}
\begin{minipage}{0.49\linewidth}
\centering
\epsfig{file=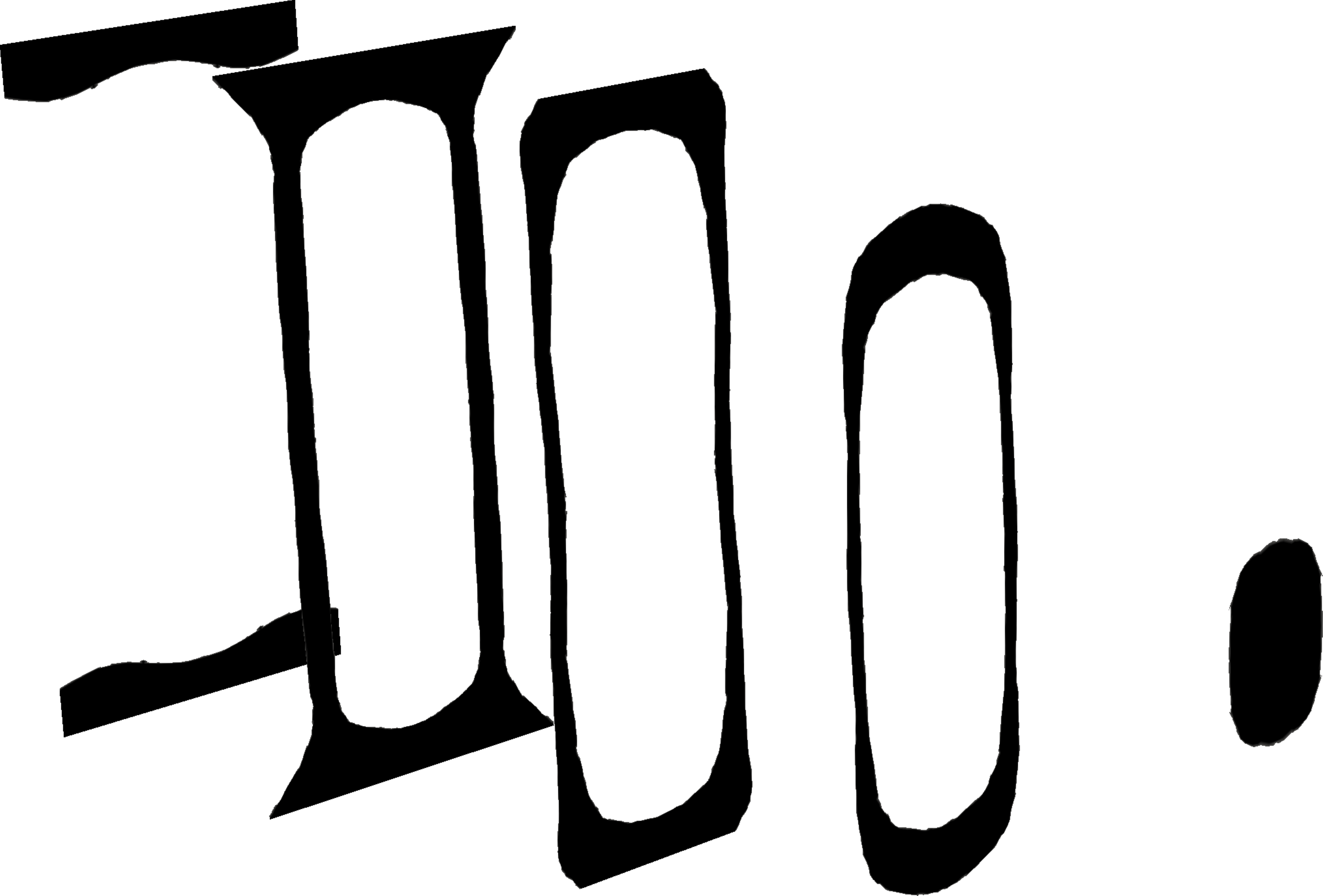, width=\linewidth}
\end{minipage}
\begin{minipage}{0.49\linewidth}
\centering
\epsfig{file=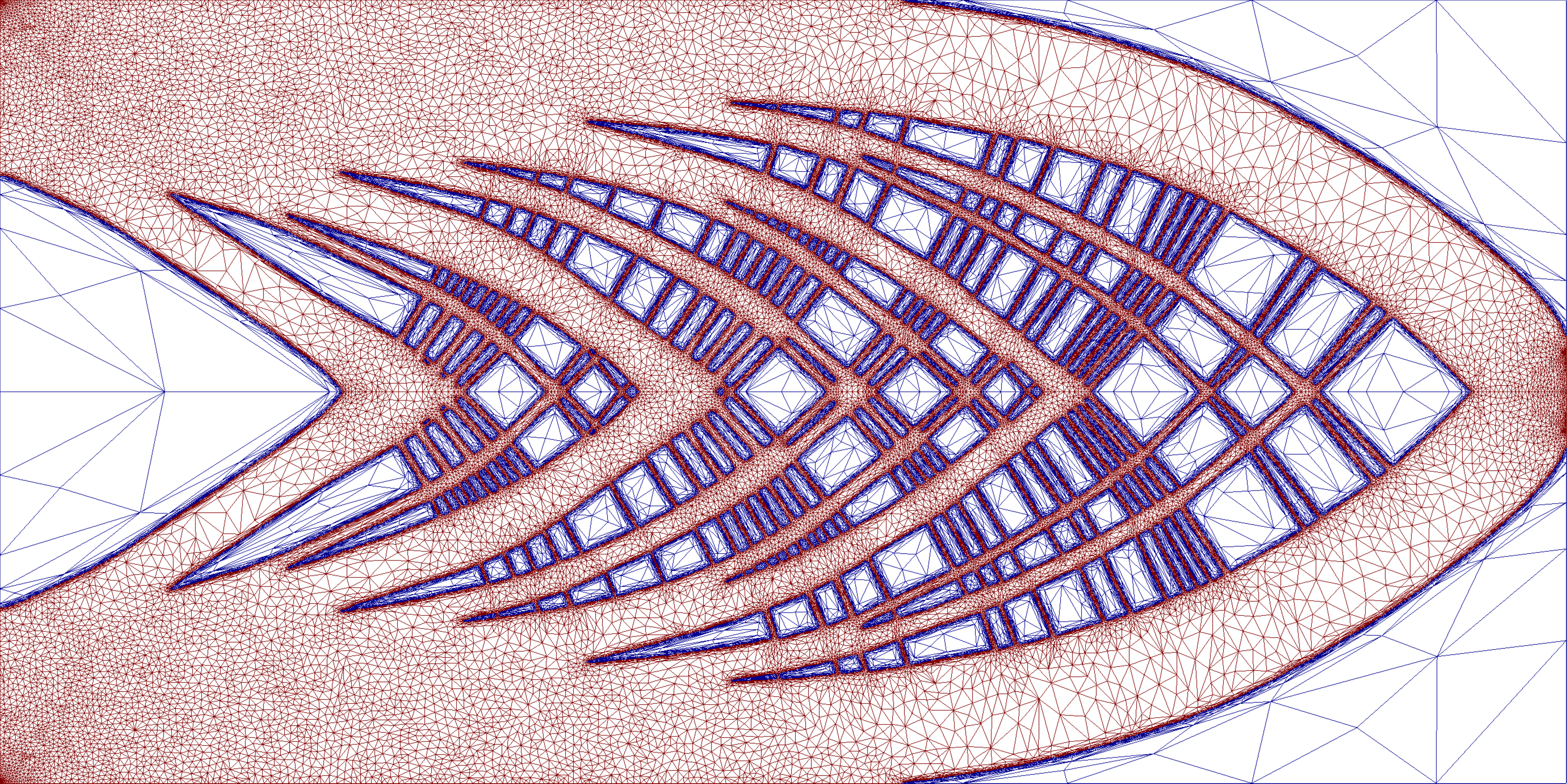, width=\linewidth}
\end{minipage}
\begin{minipage}{0.49\linewidth}
\centering
\epsfig{file=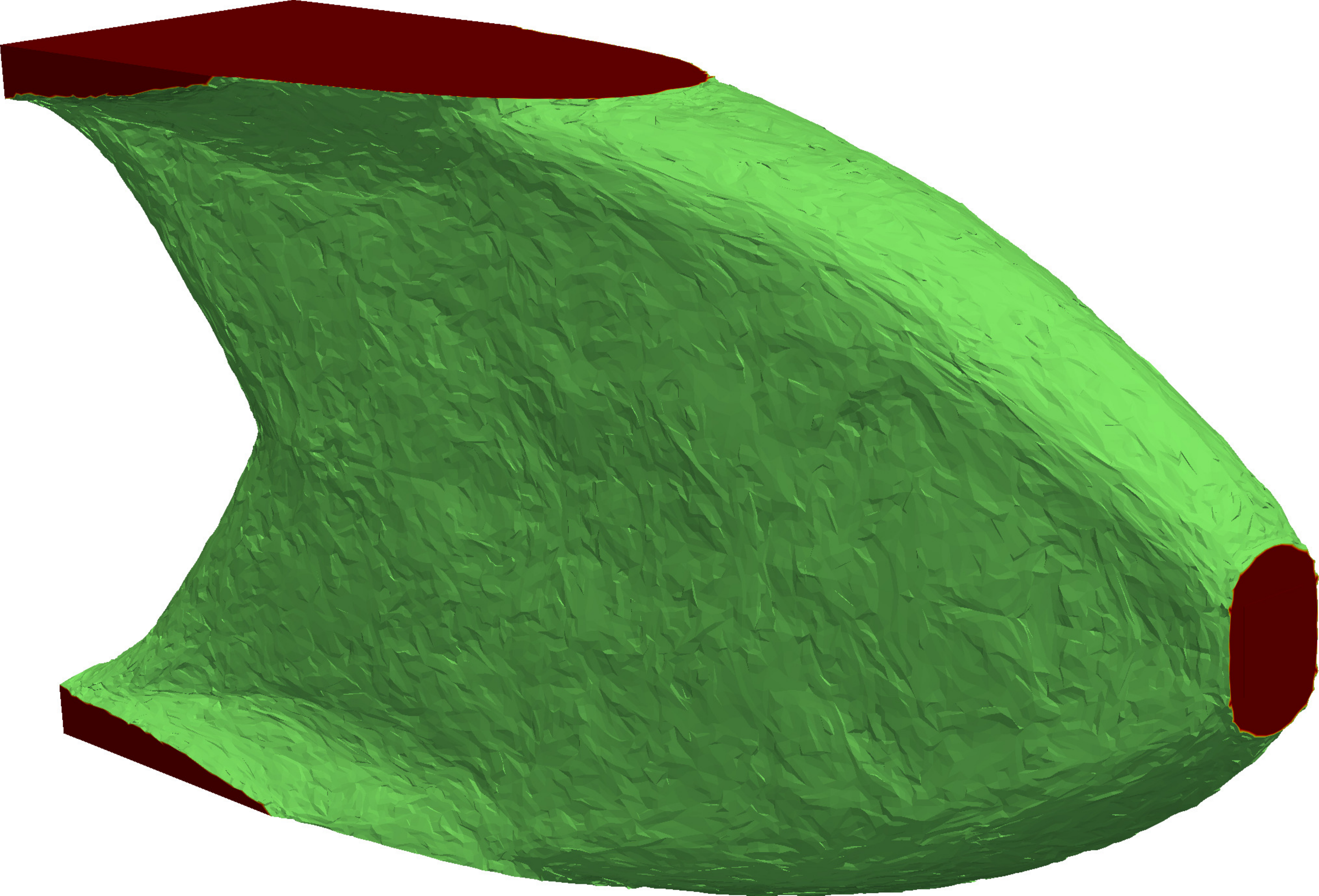, width=\linewidth}
\end{minipage}
\caption{The setup for structural compliance minimisation (top left) and the 
result for the case of a small minimum length scale (bottom left). Red 
corresponds to solid areas and blue to void. The result of compliance 
minimisation in 3D is shown in terms of the cross-sectional view (top right) 
and the solid/void interface (bottom right).}
\label{fig:cant}
\end{figure}

Structural compliance minimisation is concerned with the problem of finding 
stiff and lightweight mechanical components \cite{suresh2010199}, often in the 
context of linear elasticity. The setup for a classical cantilever problem with 
support to the left and a load to the right is shown in Figure \ref{fig:cant} 
(top left). The question is how to form the stiffest possible link between the 
two boundaries, given a certain amount of isotropic material. The problem is 
ill-posed unless a minimum length scale is imposed for the design, because the 
optimal structure is a composite. In fact, one can see a tendency towards 
microstructured areas when a small minimum length scale 
$L_\mathrm{min}=10^{-3}L_\mathrm{char}$ is used as illustrated in Figure 
\ref{fig:cant} (bottom left). Note how the many straight and parallel 
connections can be efficiently resolved with anisotropic elements. Mesh 
adaptation is driven by the Hessians of the design and the topological 
derivative \cite{suresh2010199}. A Helmholtz filter is applied to both design 
and derivative to smooth out features smaller than $L_\mathrm{min}$, before the 
Hessians are computed.

The two dimensional setup is also extruded to three dimensions, as plotted in 
Figure \ref{fig:cant} (top right and bottom right). Note that the large planar 
areas with little curvature are well resolved by the anisotropic elements. The 
increased dimensionality leads to a much simpler topology even though the 
optimisation is performed with $L_\mathrm{min}=5\cdot10^{-3}L_\mathrm{char}$.

In order to evaluate the parallel performance, a synthetic solution $\psi$ is 
defined to vary in time and space:
\begin{scriptsize}
\begin{equation*}
\psi(x,y,t) = 0.1\sin{\left(50x + \frac{2\pi t}{T}\right)} + 
  \arctan{\left(-\frac{0.1}{2x-\sin{\left(5y + \frac{2\pi t}{T}\right)}}\right)}
\end{equation*}
\end{scriptsize}
where $T$ is the period. This is a good choice as a benchmark as it contains 
multi-scale features and a shock front, \ie the typical solution 
characteristics where anisotropic adaptive mesh methods excel. An isotropic 
mesh was generated on the unit square using approximately $200\times200$ 
triangles and the adaptation benchmark was run with a requirement for $\approx 
550k$ elements. The same example was extruded in 3D, where an isotropic mesh 
was generated in the unit cube using approximately $50\times50\times50$ 
tetrahedra and the adaptation benchmark was run with a requirement for $\approx 
210k$ elements. 3D swapping has not been parallelised, therefore the 
corresponding results have been omitted.

\begin{figure}
\centering
\epsfig{file=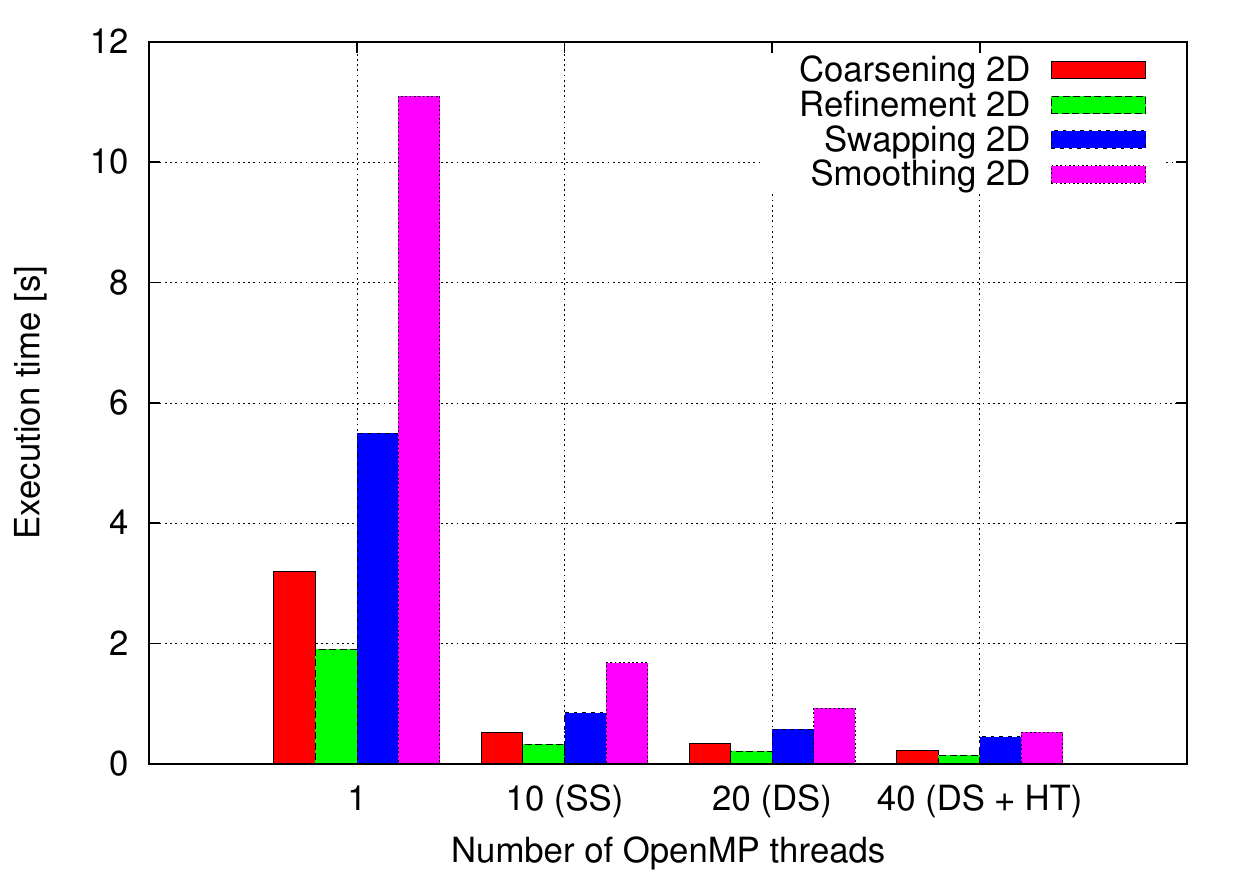, width=0.9\linewidth}
\epsfig{file=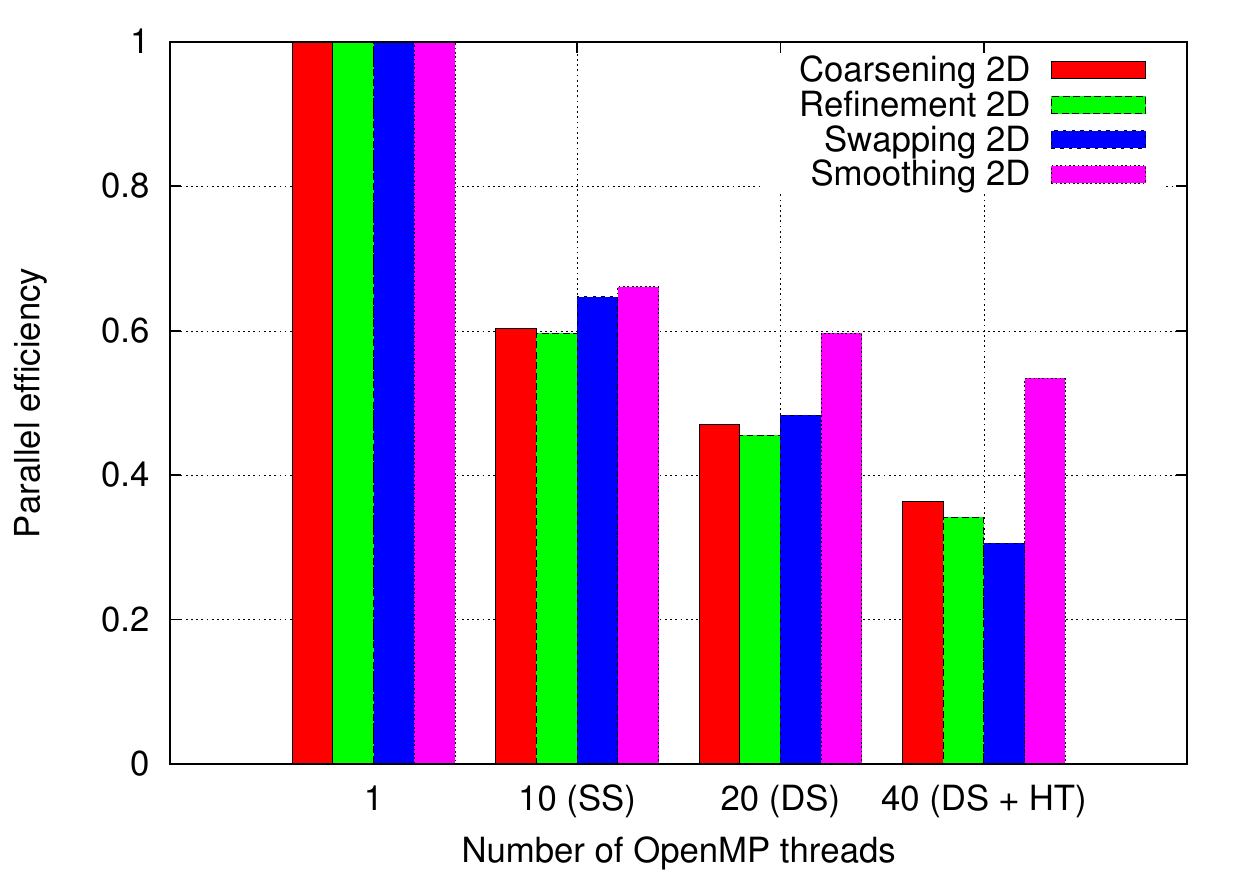, width=0.9\linewidth}
\epsfig{file=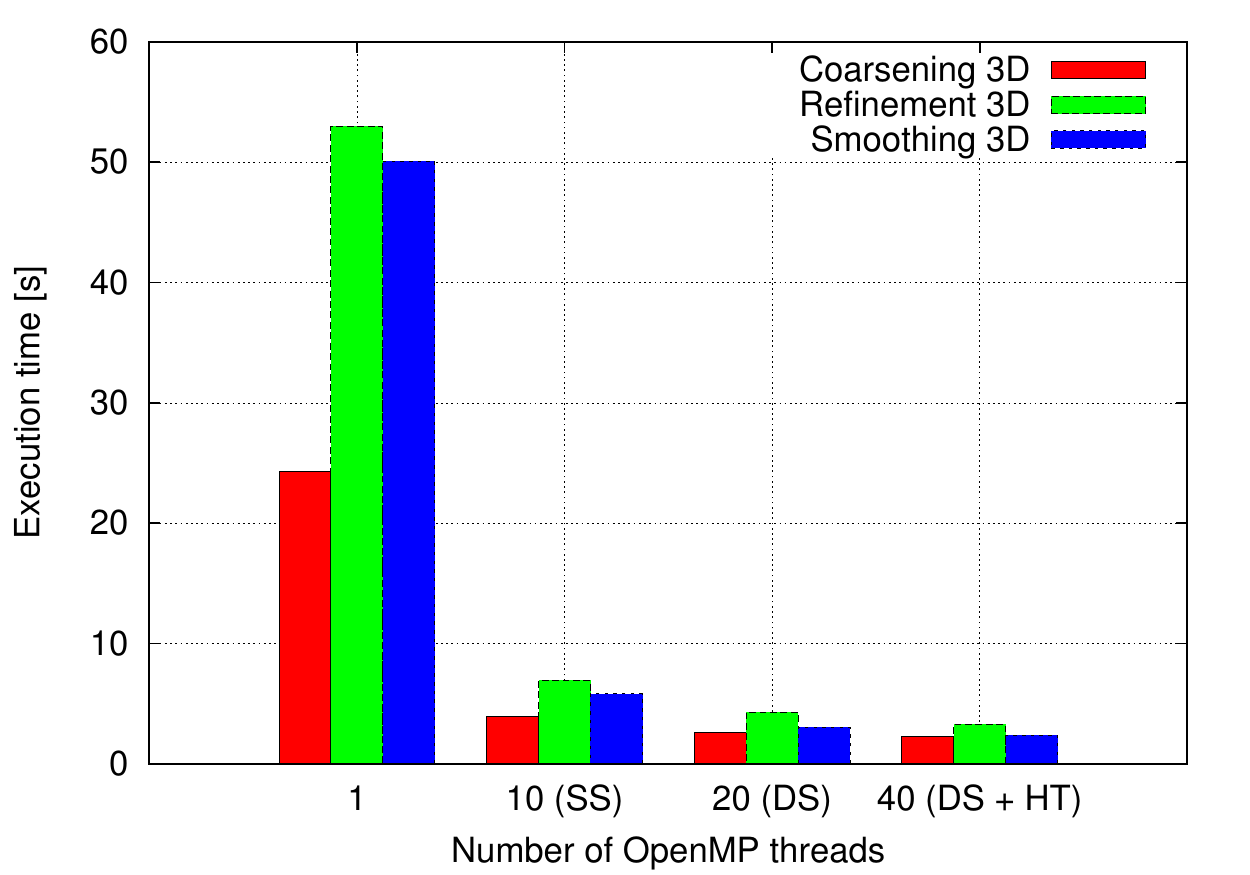, width=0.9\linewidth}
\epsfig{file=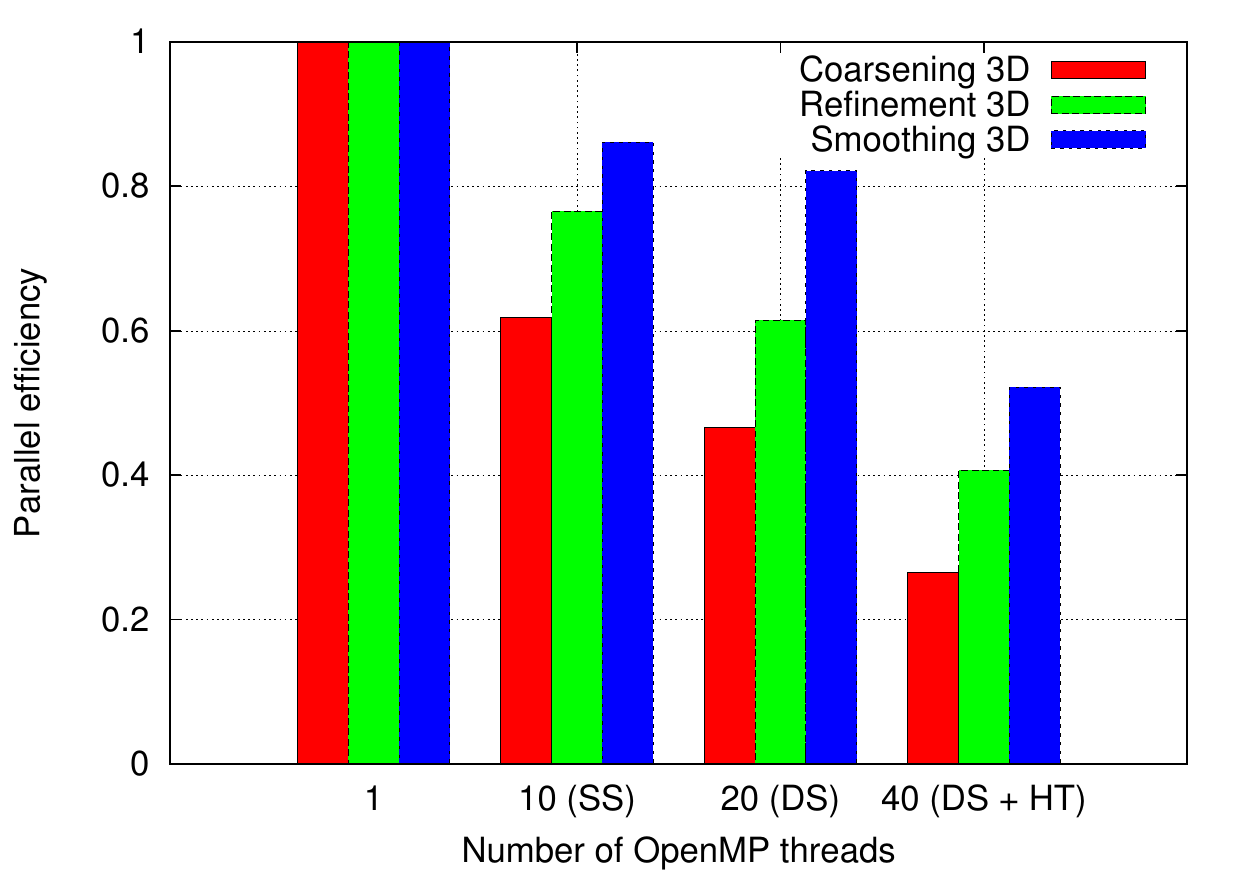, width=0.9\linewidth}
\caption{Execution time and parallel efficiency of 2D and 3D synthetic 
benchmarks on the 2x8-core \INTELXEON Sandy Bridge system.}
\label{fig:bench_intel}
\end{figure}

\begin{figure}
\centering
\epsfig{file=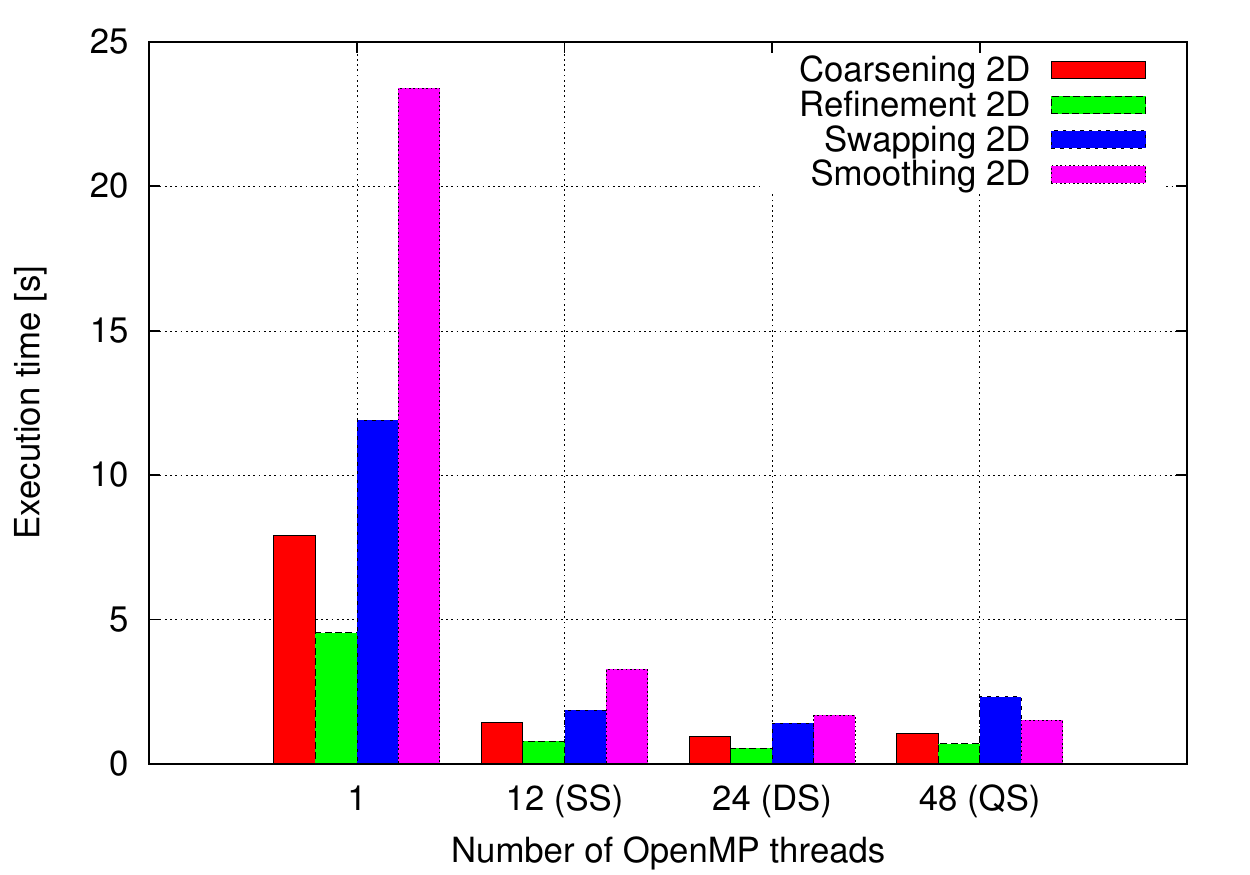, width=0.9\linewidth}
\epsfig{file=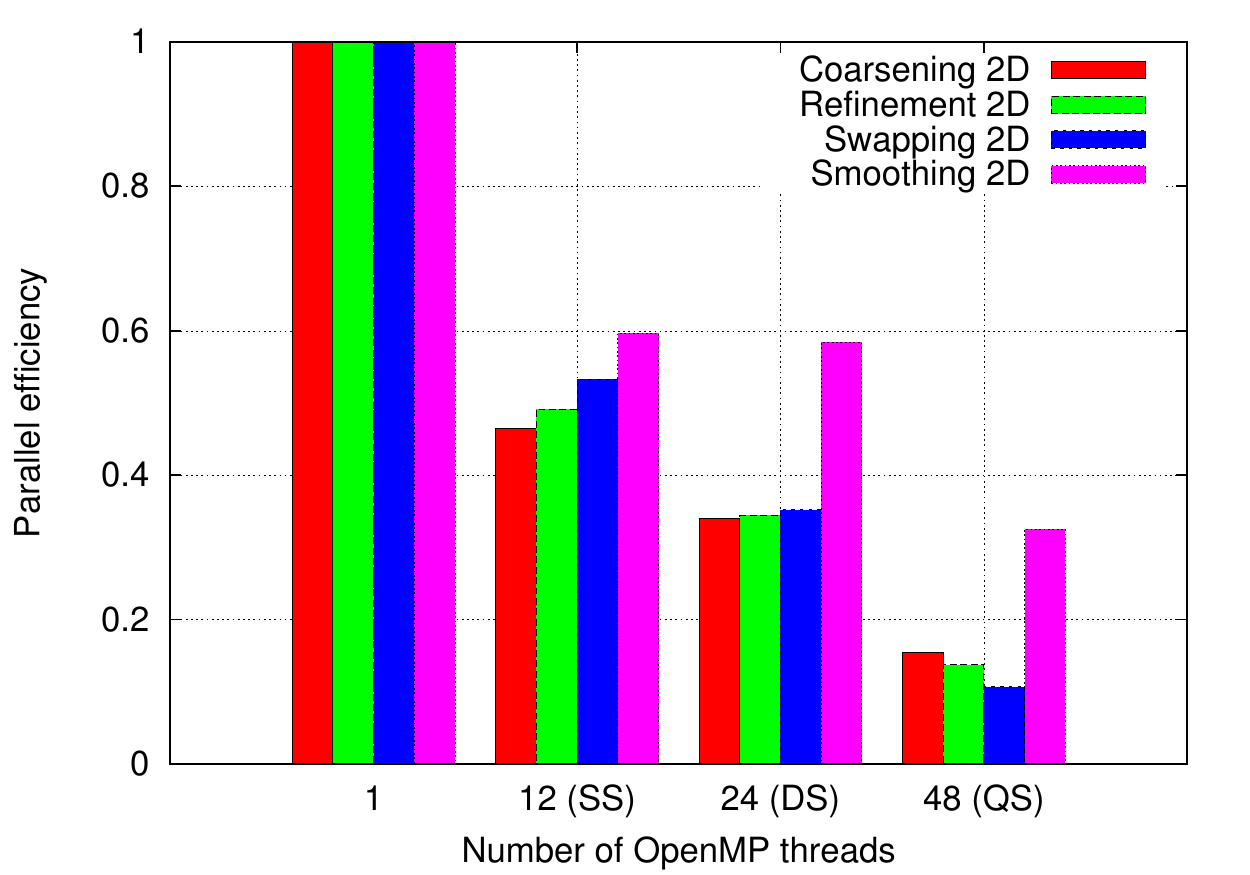, width=0.9\linewidth}
\epsfig{file=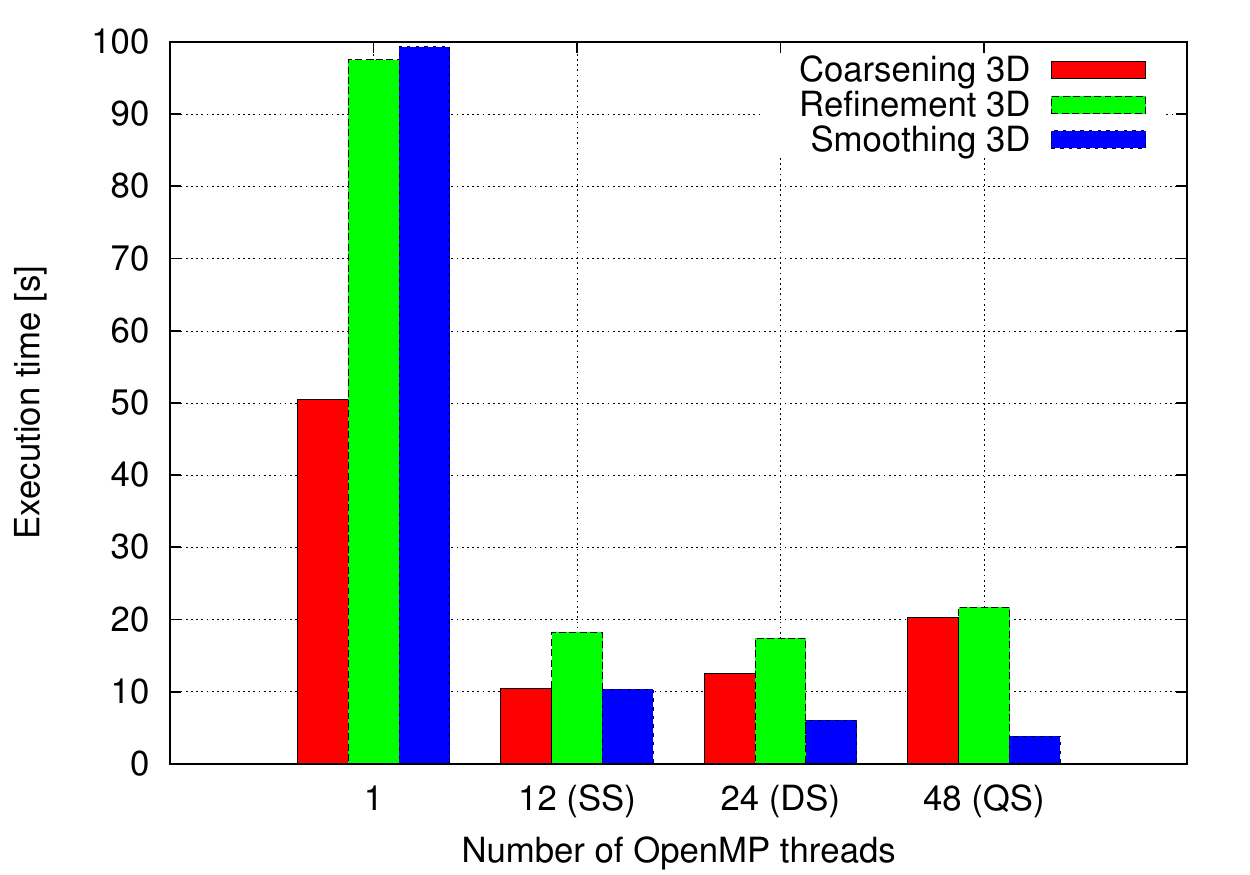, width=0.9\linewidth}
\epsfig{file=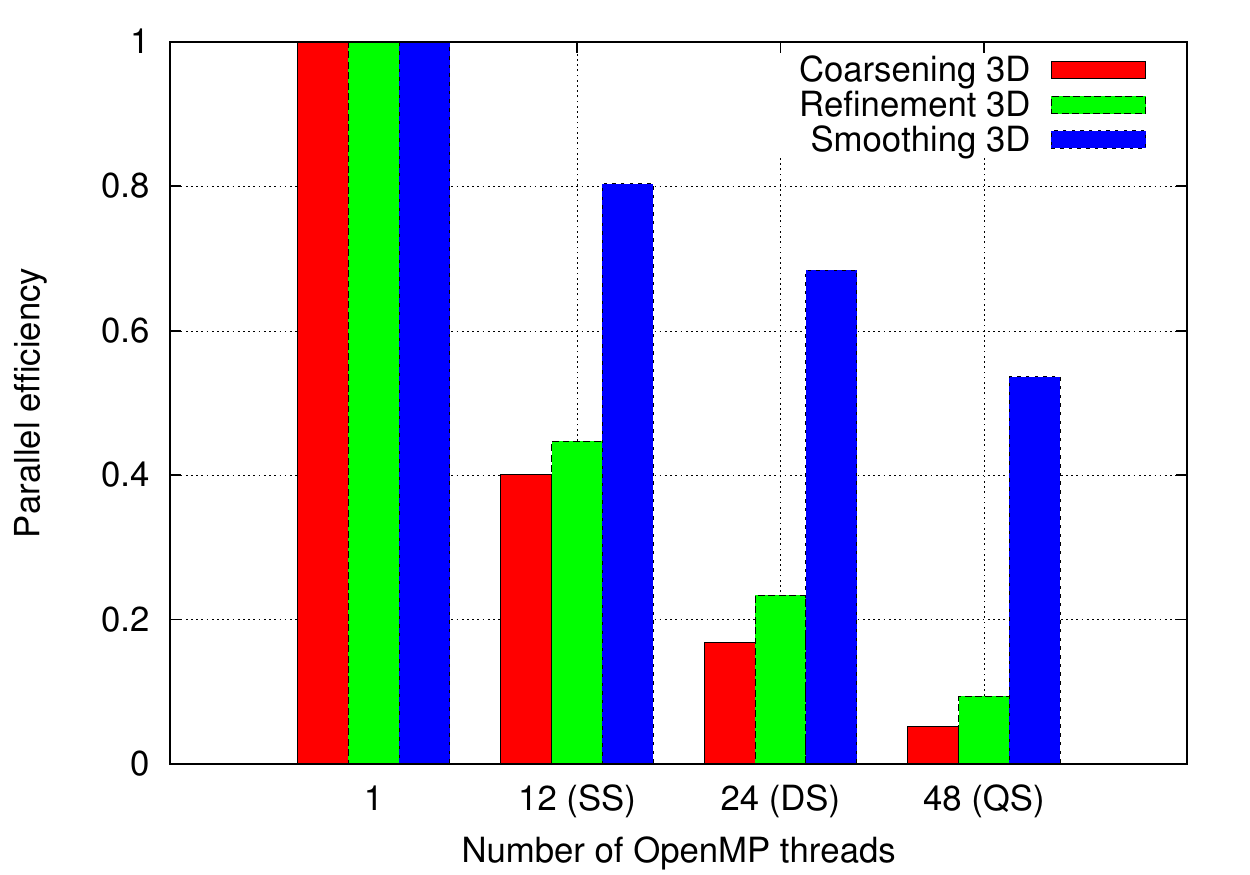, width=0.9\linewidth}
\caption{Execution time and parallel efficiency of 2D and 3D synthetic 
benchmarks on the 4x12-core \AMDOPTERON Magny-Cours system.}
\label{fig:bench_amd}
\end{figure}

The code was compiled using the Intel compiler suite (version 14.0.1) and with 
the {\tt -O3} optimisation flag. We used two systems to evaluate performance: 
(a) a dual-socket \INTELXEON E5-2650 system (Sandy Bridge, 2.00GHz, 8 cores per 
socket, 2-way hyper-threading) running \RHEL Server release 6.4 (Santiago) and 
(b) a quad-socket \AMDOPTERON 6176 system (Magny-Cours, 2.3GHz, 12 cores per 
socket) running Ubuntu 12.04.5. In all cases, thread-core affinity was used. 
Figures \ref{fig:bench_intel} and \ref{fig:bench_amd} show the average (over 50 
time steps) execution time per time step and parallel efficiency against the 
number of threads using single-socket (SS), dual-socket (DS) and quad-socket 
(QS) configurations. On the \INTELXEON system we also enable hyper-threading 
(HT) to make use of all 40 logical cores.

Running on one socket, our code achieves a parallel efficiency of over 60\% on 
\INTELXEON and around 50\% on \AMDOPTERON. Smoothing scales better than the 
other algorithms as it is more compute-intensive, which favours scalability, 
reaching an efficiency of over 50\% even in the quad-socket case. When we move 
to more sockets, NUMA effects become pronounced, which is expected as common 
NUMA optimisations such as pinning and first-touch for page binding are 
ineffective for irregular computations. Nonetheless, the achievable efficiency 
is good considering the irregular nature of those algorithms.

\section{Conclusion}
\label{sect:conclusion}
In this paper we examined the scalability of anisotropic mesh adaptivity using 
a thread-parallel programming model and explored new parallel algorithmic 
approaches to support this model. Despite the complex data dependencies and 
inherent load imbalances we have shown it is possible to achieve practical 
levels of scaling using a combination of a fast graph colouring technique, the 
deferred-update strategy, atomic-based creation of worklists and for-loop 
work-stealing. In principle, this methodology facilitates scaling up to the 
point where the number of elements of an independent set is equal to the number 
of available threads.

\section*{Acknowledgments}
This project has been supported by Fujitsu Laboratories of Europe Ltd and the 
EPSRC (grant numbers EP/I00677X/1 and EP/L000407/1).

\bibliographystyle{abbrv}
\bibliography{article}

\balancecolumns

\end{document}